\documentclass[preprint,12pt]{elsarticle}
\usepackage{graphicx}
\usepackage{amssymb}
\usepackage{amsmath}
\usepackage{subfig}
\usepackage{multirow}
\usepackage{colortbl,color}
\usepackage{algorithm}
\usepackage{amsthm}
\usepackage{hyperref}
\biboptions{comma,sort&compress}
\journal{ }

\begin{document}

\begin{frontmatter}
\title{Multi-step Steady-State Measurements of Low Permeability Using Series Circuit with A Reference Rock Sample}
\author[KFUPM]{Jun~Li\footnote[1]{e-mail: lijun04@gmail.com, \href{https://www.researchgate.net/profile/Jun_Li99}{URL}.}}
\address[KFUPM]{Center for Integrative Petroleum Research, \\ College of Petroleum Engineering and Geosciences, \\ King Fahd University of Petroleum $\&$ Minerals, Saudi Arabia}
\begin{abstract}
A multi-step steady-state (MSSS) method is proposed here for the measurement of low permeability. This new method can accurately and easily measure very low permeabilities of rock samples using a new setup, where the targeted rock sample and ordinary apparatus components are connected with a reference rock sample to form a series circuit. Any conventional rock sample with high permeability could be used as a reference rock sample such that the traditional steady-state measurement is feasible to accurately determine its permeability as a reference value in the MSSS method. The challenging measurement of tiny mass flux rate by advanced pump system is avoided and the permeability of targeted rock sample can be directly computed using the pressure drops, sectional areas and lengths of the two connected rock samples, and the known permeability of the reference rock sample, based on the mass conservation principle in a series circuit at steady state. Multi-step measurements using additional reference rock samples will be needed if the pressure drop across the first reference rock sample is too small to be accurately measured due to high permeability ratio when it is connected with the targeted rock sample to form a series circuit. The relative pressure drops can be small since the measurement of flow speed is unnecessary, which improves the accuracy in studying the dependence of gas permeability on the pore pressure. Consequently, the advantages of the MSSS method include low expense, simplicity, high accuracy and efficiency. 
\end{abstract}
\end{frontmatter}
\section{Introduction}\label{s:intro}
The recently developed pulse-decay scheme is actually not purely based on measurements but also depends on the accuracy/validity of an empirical partial-differential equation (PDE), which is used to interpret the measured pressure evolution at the outlet by regulating the permeability contained in the PDE as a parameter until the pressure evolution at the outlet, which is computed using the PDE, the permeability parameter and the given pressure pulse at the inlet, matches the measured pressure evolution at the outlet \cite{Jin2015}.

The traditional experimental techniques of measuring permeability at steady state \cite{Sinha2013}-\cite{Katsuki2016} are very time-consuming when the permeability of the rock sample is very low making the measurement of tiny mass flux rate through the rock sample difficult. In order to facilitate the measurement by increasing the flux rate to reduce noise, the pressure difference between the inlet and outlet of the rock sample needs to be large, which unfortunately leads to upscaling error when the objective of measurement is to obtain the dependence of gas permeability on the specific pore pressure, instead of the average pore pressure over a large pressure difference between the two ends.

\section{Multi-step Steady-State Method for Low Permeability Measurement}\label{s:multistep}
In the new steady-state measurement proposed here, we use the same principle, theory and mechanism of traditional steady-state measurement but completely change the setup of apparatus components such that the measurement of tiny mass flux rate becomes unnecessary, which is achieved by connecting a reference rock sample of known permeability with the targeted rock sample to form a series circuit that connects at its two ends to two gas/liquid cylinders with different but fixed pressures. The two cylinders should be large enough to maintain a quasi steady state for a while. This scheme is applicable to the permeability measurements for both gas and liquid through tight or shale rock samples. Since the mass flux rates through the two connected rock samples should be the same at steady state according to the mass conservation law, the permeability of the targeted rock sample can be analytically computed using the pressure drops, sectional areas and lengths of the two rock samples, and the known permeability of the reference rock sample. Any conventional rock sample with high permeability could be used as a reference rock sample such that the traditional steady-state measurement is feasible to accurately determine its permeability as a reference value in the new measurement method. If gas permeability is under study, accurate measurement of the dependence of permeability on the pore pressure requires that the relative pressure difference across the targeted rock sample should be as small as possible. Consequently, the total pressure drop between the two gas cylinders should be as small as possible but also needs to be much larger than the accuracy of differential pressure transducer used, which can be easily satisfied in experiments. Although the sum of pressure drops across the two rock samples is equal to the preset total pressure drop between the two cylinders, the reference rock sample with higher permeability usually has much smaller pressure drop than the targeted rock sample with lower permeability if their sectional-area ratio and length ratio are close to 1. When the pressure drop across the reference rock sample is too small to be accurately measured using the available differential pressure transducers due to very high permeability ratio, multi-step steady-state measurements will be used to reduce the ratio of two pressure drops in each step. 

For example, we have reference sample $a$ of 100 mD (known), sample $b$ of 10 mD, targeted sample $c$ of 1 mD. We want to use a pressure drop of only $10^4$ Pa such that the relative pressure variation is negligible, and assume that the value below $10^2$ Pa given by the differential pressure transducers is not reliable. If we connect the samples $a$ and $c$ together, the pressure drop across the reference sample $a$ is about 99 Pa, which cannot be accurately measured. But, if we use the samples $a$ and $b$ together as the \textit{first} step, the pressure drops across each of them will be 909 and 9091 Pa, respectively, which can be accurately measured to first determine the permeability of the sample $b$. Then, the permeability of the targeted sample $c$ can be similarly determined by connecting the samples $b$ and $c$ together in the \textit{second} step. The specific formulas used in the permeability calculation will be given below.

In the multi-step steady-state measurements, an auxiliary rock sample $\# 1$ that has permeability lower than the reference rock sample (i.e., reference rock sample $\# 1$ with known permeability) but higher than the targeted rock sample is used together with the reference rock sample $\# 1$ to form a series circuit for the first-step measurement, where the permeability of the auxiliary rock sample $\# 1$ can be determined as discussed above and shown in Eqs.~\eqref{eq1}-\eqref{eq3} in general. Then, in the second-step measurement, the auxiliary rock sample $\# 1$ is taken as the reference rock sample $\# 2$ and {\textit{tentatively}} connected with the targeted rock sample to form a series circuit. Similarly, if the pressure drop across the reference rock sample $\# 2$ is too small to be accurately measured due to very high permeability ratio between the two connected rock samples, another auxiliary rock sample $\# 2$ that has permeability lower than the reference rock sample $\# 2$ but higher than the targeted rock sample is used together with the reference rock sample $\# 2$ to form a series circuit for the second-step measurement, where the permeability of the auxiliary rock sample $\# 2$ can be determined. Multi-steps might be repeated till the pressure drop across the last reference rock sample $\# n$ can be accurately measured when the reference rock sample $\# n$ is connected with the targeted rock sample to form a series circuit and then, the permeability of the targeted rock sample can be eventually determined. We note that the total step number can be significantly reduced by using differential pressure transducer with high accuracy.

\begin{figure}[H]
\centering
  \includegraphics[width=0.95\textwidth]{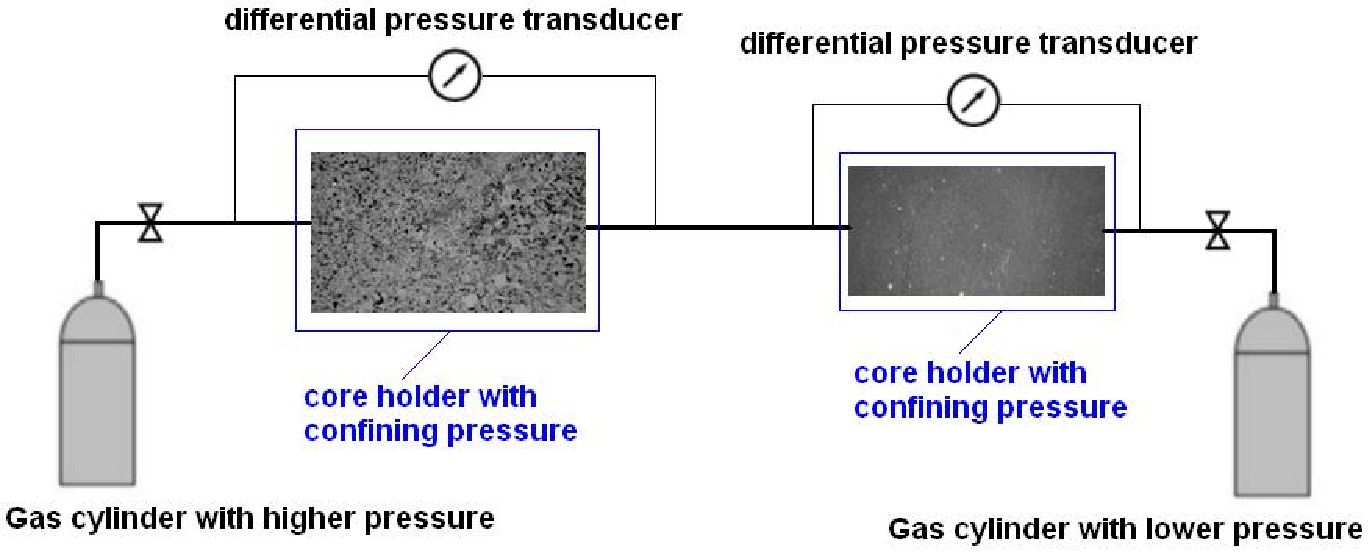}
\caption{Schematic of low-permeability measurement using series circuit with a reference rock sample.}
\label{fig:schematic}
\end{figure}

The measurement procedure for each step (e.g., the step $\# n$ in general) is discussed as follows. Experimental setup is given in Fig.~\ref{fig:schematic} as an example, where the left rock sample (i.e., reference rock sample $\# n$) has higher permeability than the right one (i.e., auxiliary rock sample $\# n$ or the targeted rock sample). We denote the pressure drop, sectional area, rock length in the flow direction and permeability of the reference rock sample $\# n$ by $\Delta p_{n,\rm{ref}}$,  $A_{n,\rm{ref}}$,  $L_{n,\rm{ref}}$, and  $\kappa_{n,\rm{ref}}$, respectively. Similarly, we have  $\Delta p_{n,\rm{aux/tar}}$,  $A_{n,\rm{aux/tar}}$,  $L_{n,\rm{aux/tar}}$, and  $\kappa_{n,\rm{aux/tar}}$ for the auxiliary rock sample $\# n$ or the targeted rock sample. The average fluid densities $\rho$ in the two connected rock samples are almost the same due to relatively small total pressure drop between the two cylinders. After opening the valves, the flow is deemed to be at steady state when the variations of $\Delta p_{n,\rm{ref}}$ and $\Delta p_{n,\rm{aux/tar}}$ are negligible and this process should be very quick. Then, according to the \textit{definition} of permeability and the facts that the two connected rock samples have the same mass flux rate $Q$ at steady state and the dynamic viscosity $\mu$ of fluid is almost the same in the two rock samples with relatively small pressure difference, we have:
\begin{equation}\label{eq1}
    \kappa_{n,\rm{ref}}=\dfrac{\mu QL_{n,\rm{ref}}}{\rho A_{n,\rm{ref}}\Delta p_{n,\rm{ref}}},
\end{equation}
and
\begin{equation}\label{eq2}
    \kappa_{n,\rm{aux/tar}}=\dfrac{\mu QL_{n,\rm{aux/tar}}}{\rho A_{n,\rm{aux/tar}}\Delta p_{n,\rm{aux/tar}}}.
\end{equation}

Combining Eqs.~\eqref{eq1} and \eqref{eq2}, we get the formula to compute $\kappa_{n,\rm{aux/tar}}$ as follows:
\begin{equation}\label{eq3}
    \kappa_{n,\rm{aux/tar}}=\kappa_{n,\rm{ref}}\dfrac{A_{n,\rm{ref}}\Delta p_{n,\rm{ref}}L_{n,\rm{aux/tar}}}{A_{n,\rm{aux/tar}}\Delta p_{n,\rm{aux/tar}}L_{n,\rm{ref}}},
\end{equation}
where $\kappa_{n,\rm{ref}}$ is originally known for the first reference rock sample or becomes known for arbitrary reference rock sample $\# n$, i.e. auxiliary rock sample $\# (n-1)$, during the measurement of previous step $\# (n-1)$. Again, in the above example of using two steps for three samples $a, b, c$, the general Eq.~\eqref{eq3} takes the following form: 
\begin{equation}\label{eq4}
\begin{split}
    \kappa_{2,\rm{tar}}&=\kappa_{2,\rm{ref}}\dfrac{A_{2,\rm{ref}}\Delta p_{2,\rm{ref}}L_{2,\rm{tar}}}{A_{2,\rm{tar}}\Delta p_{2,\rm{tar}}L_{2,\rm{ref}}}\\
                               &=\kappa_{1,\rm{aux}}\dfrac{A_{2,\rm{ref}}\Delta p_{2,\rm{ref}}L_{2,\rm{tar}}}{A_{2,\rm{tar}}\Delta p_{2,\rm{tar}}L_{2,\rm{ref}}}\\
                               &=\kappa_{1,\rm{ref}}\dfrac{A_{1,\rm{ref}}\Delta p_{1,\rm{ref}}L_{1,\rm{aux}}}{A_{1,\rm{aux}}\Delta p_{1,\rm{aux}}L_{1,\rm{ref}}} \dfrac{A_{2,\rm{ref}}\Delta p_{2,\rm{ref}}L_{2,\rm{tar}}}{A_{2,\rm{tar}}\Delta p_{2,\rm{tar}}L_{2,\rm{ref}}},                    
\end{split}                               
\end{equation}
where $\kappa_{2,\rm{tar}}=\kappa_c$, $\kappa_{1,\rm{ref}}=\kappa_a$, $A_{1,\rm{ref}}=A_a$, $\Delta p_{1,\rm{ref}}=(\Delta p_a)_{\rm step 1}$, $L_{1,\rm{aux}}=L_b$, $A_{1,\rm{aux}}=A_b$, $\Delta p_{1,\rm{aux}}=(\Delta p_b)_{\rm step 1}$, $L_{1,\rm{ref}}=L_a$, $A_{2,\rm{ref}}=A_b$, $\Delta p_{2,\rm{ref}}=(\Delta p_b)_{\rm step 2}$, $L_{2,\rm{tar}}=L_c$, $A_{2,\rm{tar}}=A_c$, $\Delta p_{2,\rm{tar}}=(\Delta p_c)_{\rm step 2}$, $L_{2,\rm{ref}}=L_b$. Thus, Eq.~\eqref{eq4} can be rewritten into: 
\begin{equation}\label{eq5}
\begin{split}
    \kappa_c&=\kappa_a\dfrac{A_a(\Delta p_a)_{\rm step 1}L_b}{A_b(\Delta p_b)_{\rm step 1}L_a}\dfrac{A_b(\Delta p_b)_{\rm step 2}L_c}{A_c(\Delta p_c)_{\rm step 2}L_b}\\
                   &=\kappa_a\dfrac{A_a(\Delta p_a)_{\rm step 1}}{(\Delta p_b)_{\rm step 1}L_a}\dfrac{(\Delta p_b)_{\rm step 2}L_c}{A_c(\Delta p_c)_{\rm step 2}}. 
\end{split}      
\end{equation}

We should note that the permeability determinations of auxiliary rock samples by the multi-step steady-state measurements can be done once for all and thus it is advisable to preserve a benchmarked auxiliary/reference rock sample set that should contain different rock samples with permeability ranging from Darcy, milli-Darcy, micro-Darcy to nano-Darcy, for instance, to be used repeatedly.  For the gas permeability measurement with slippage effect, each auxiliary/reference rock sample has different permeability values at different pressures (i.e., pore pressures), temperatures or confining pressures that change the pore space and so the right value at the pore pressure, temperature and confining pressure concerned should be used in Eq.~\eqref{eq3}. Additionally, the type of gas media also has influence on the permeability and thus the permeability calibration of the auxiliary/reference rock sample set should be done for each type of gas media of interest as well.

\section{Conclusions}\label{s:conclusions}
A novel method (i.e., MSSS method) using multi-step steady-state measurements is proposed to measure the low gas/liquid permeability in unconventional rock sample. The challenging measurement of tiny mass flux is avoided in the MSSS method based on the mass conservation principle in a series circuit with a reference rock sample. The MSSS method is simpler, cheaper, more accurate and efficient than the traditional methods for steady-state measurements, and robust for different sample sizes. The features of the MSSS method are summarized as follows:

The principle, theory and mechanism of the MSSS method are exactly the same as those of the traditional steady-state measurements and thus the validity of the MSSS method is guaranteed. Compared to the traditional methods, the MSSS method allows using smaller pressure difference across the targeted rock sample and thus is more accurate in studying the dependence of permeability on the pore pressure for gas flow problems. 

The novelty of the MSSS method lies in a new arrangement of ordinary apparatus components that are commonly used in the traditional steady-state measurements. With the help of a reference rock sample, this novelty can avoid the measurement of tiny mass flux inside the targeted rock sample and consequently, some traditional components (e.g., advanced and expensive pumps) are not needed by the MSSS method and thus the cost can be significantly reduced. 

The traditional steady-state measurements actually take two stages to complete: a) open valves and advance piston at constant speed to reach steady state; b) measure the mass flux after reaching steady state, which is usually the time-consuming part since the movement of piston is very slow. In contrast, the MSSS method only needs the first stage to reach steady state and thus is certainly more efficient than the traditional steady-state measurements although the exact improvement of efficiency will be problem-dependent.  
%


\begin{thebibliography}{23}
\expandafter\ifx\csname natexlab\endcsname\relax\def\natexlab#1{#1}\fi
\providecommand{\bibinfo}[2]{#2}
\ifx\xfnm\relax \def\xfnm[#1]{\unskip,\space#1}\fi

\bibitem{Jin2015}
\bibinfo{author}{G.D.~Jin}, \bibinfo{author}{P.H.~Gonzalez}, \bibinfo{author}{A.A.~Al-Dhamen}, \bibinfo{author}{S.S.~Ali}, \bibinfo{author}{A.~Nair}, \bibinfo{author}{G.~Agrawal}, \bibinfo{author}{M.R.~Khodja}, \bibinfo{author}{S.R.~Hussaini}, \bibinfo{author}{Z.Z.~Jangda}, and \bibinfo{author}{A.Z.~Ali},  \newblock \bibinfo{title}{Permeability Measurement of Organic-rich Shale - Comparison of Various Unsteady-state Methods},
\bibinfo{year}{2015, SPE-175105}.

\bibitem{Sinha2013}
\bibinfo{author}{S.~Sinha}, \bibinfo{author}{E.M.~Braun}, \bibinfo{author}{M.D.~Determan}, \bibinfo{author}{Q.R.~Passey}, \bibinfo{author}{S.A.~Leonardi}, \bibinfo{author}{J.A.~Boros}, \bibinfo{author}{A.C.~Wood III}, \bibinfo{author}{T.~Zirkle}, and \bibinfo{author}{R.A.~Kudva},
\newblock \bibinfo{title}{Steady-State Permeability Measurements on Intact Shale Samples at Reservoir Conditions - Effect of Stress, Temperature, Pressure, and Type of Gas},
\bibinfo{year}{2013, SPE-164263}.

\bibitem{Zamirian2014}
\bibinfo{author}{M.~Zamirian}, \bibinfo{author}{K.~Aminian}, \bibinfo{author}{S.~Ameri}, and \bibinfo{author}{E.~Fathi},
\newblock \bibinfo{title}{New Steady-State Technique for Measuring Shale Core Plug Permeability},
\bibinfo{year}{2014, SPE-171613}.

\bibitem{Katsuki2016}
\bibinfo{author}{D.~Katsuki}, \bibinfo{author}{A.P.~Deben}, \bibinfo{author}{O.~Adekunle}, \bibinfo{author}{A.J.~Rixon}, and \bibinfo{author}{A.N.~Tutuncu},
\newblock \bibinfo{title}{Stress-Dependent Permeability and Dynamic Elastic Moduli of Reservoir and Seal Shale},
\bibinfo{year}{2016, URTeC-2461613}.


\end{thebibliography}
\end{document}